\documentclass[11pt]{revtex4}
\usepackage{graphicx}
\usepackage{amssymb}

\begin{document}

\title[Visualizing spreading surfactants]{Fluorescent visualization of a spreading surfactant}

\author{David W. Fallest,$^1$ Adele M. Lichtenberger,$^1$ Christopher J. Fox$^{1,2}$ and Karen E. Daniels$^1$\footnote{kdaniel@ncsu.edu}}

\address{$^1$ Department of Physics, North Carolina State University, Raleigh, NC, 27695}
\address{$^2$ Department of Mathematics, Harvey Mudd College, Claremont, CA}

\begin{abstract}
The spreading of surfactants on thin films is an industrially and medically important phenomenon, but the dynamics are highly nonlinear and visualization of the surfactant dynamics has been a long-standing experimental challenge. We perform the first quantitative, spatiotemporally-resolved measurements of the spreading of an insoluble surfactant on a thin fluid layer. During the spreading process, we directly observe both the radial height profile of the spreading droplet and the spatial distribution of the fluorescently-tagged surfactant. We find that the leading edge of spreading circular layer of surfactant forms a Marangoni ridge in the underlying fluid, with a trough trailing the ridge as expected. However, several novel features are observed using the fluorescence technique, including a peak in the surfactant concentration which trails the leading edge, and a flat, monolayer-scale spreading film which differs from concentration profiles predicted by current models.  Both the Marangoni ridge and surfactant leading edge can be described to spread as $R \propto t^{\delta}$. We find spreading exponents, $\delta_H \approx 0.30$ and $\delta_\Gamma \approx 0.22$ for the ridge peak and surfactant leading edge, respectively, which are in good agreement with theoretical predictions of $\delta = 1/4$. In addition, we observe that the surfactant leading edge initially leads the peak of the Marangoni ridge, with the peak later catching up to the leading edge. 
\end{abstract}

\maketitle

\section{Introduction}

The spreading of surfactant over a thin layer of fluid is of fundamental importance to diverse biomedical applications such as pulmonary surfactant replacement therapy \cite{Enhorning-1989-SRA, Corbet-1991-DMR}, lipid tear layers in our eyes \cite{McCulley-2004-LLT}, and drug delivery mechanisms \cite{Haitsma-2001-ESD}, as well as many industrial coating and draining flows. The dynamics of the spreading process are driven by highly nonlinear equations modeling the curvature of the fluid's free surface and gradients in the surfactant concentration \cite{Borgas-1988-MFT, Gaver-1990-DLS, Troian-1990-MFI, Jensen-1992-ISS, Shen-1994-EIC}. Progress in assessing the efficacy of these theoretical models, recently reviewed by Craster and Matar \cite{Craster-2009-DST, Matar-2009-DSS}, has been hampered by the difficulty of directly visualizing the location of the surfactant in experiments \cite{Gaver-1992-DST, Bull-1999-SSS, Dussaud-2005-SCI}.

An experimentally convenient and commonly modeled geometry starts from a radially symmetric droplet of surfactant which is allowed to spread over the underlying fluid \cite{Borgas-1988-MFT, Jensen-1992-ISS, Ahmad-1972-SQT, Gaver-1992-DST, Dussaud-2005-SCI, Bull-1999-SSS}. The gradient in surfactant concentration at the leading edge of the droplet causes a gradient in surface tension. As a result, Marangoni forces pull outward on both the surfactant and the underlying fluid, causing the droplet to spread. During the spreading, a fluid ridge known as a Marangoni ridge forms near the leading edge of the surfactant. 
In models coupling lubrication theory for a thin fluid to the surface tension gradient provided by the surfactant concentration gradient, the Marangoni ridge and the peak or edge of the surfactant concentration profile have been associated with shocks or traveling wave solutions \cite{Jensen-1992-ISS, Jensen-1994-SSS, Levy-2006-MTL, Levy-2007-GDT}. However, few experimental measurements have been made to quantify the relationship between the surfactant concentration and the fluid surface curvature, due to the difficulty of directly measuring the spatial distribution of surfactant molecules \cite{Bull-1999-SSS,  Dussaud-2005-SCI}.

Previous experiments have provided quantitative measurements of either the position of the surfactant leading edge or of the fluid surface profile, but not quantitative profiles of both for the same system. Bull et al.\ \cite{Bull-1999-SSS} measured the location of the leading edge of a spreading fluorescent surfactant on a glycerin layer, while fluorescent microspheres on the surface were used to infer the arrival of the surface disturbance. Dussaud et al.\ \cite{Dussaud-2005-SCI}, measured changes in the surface slope due to the spread of oleic acid on a glycerin-water layer via Moir\'{e} topography and used talc particles on the surface to infer the arrival of the leading edge of the surfactant. In both studies, it was observed that the spreading surfactant lagged behind the surface disturbance, which we are calling the Marangoni ridge. In the experiments presented here, we provide simultaneous measurements of both the surface height profile and the surfactant concentration field by combining laser profilometry with further development of the fluorescence techniques introduced by \cite{Bull-1999-SSS}. This combination of techniques allows us to directly measure the spatiotemporal evolution of a droplet of a fluorescently-tagged surfactant lipid spreading on a thin layer of glycerin without introducing additional contaminants to the system. We find that, at later times, while the forward part of the surface disturbance can be said to be in front of the leading edge of the spreading surfactant, the surfactant leading edge coincides with the peak of the Marangoni ridge.

We compare our experimental results with the models of Jensen and Grotberg \cite{Jensen-1992-ISS, Jensen-1994-SSS} which predict a $R \sim t^{1/4}$ spreading law for the peak of the Marangoni ridge, similar to what was previously observed in experiments by Gaver and Grotberg \cite{Gaver-1992-DST}. (Note that this behavior is {\it slower} than that of droplets spreading on thick layers, for which the scaling is  $t^{3/8}$ to $t^{3/4}$, depending on the degree of surface contamination \cite{Fay-1969, Hoult-1972-OSS, Jensen-1995-SIS} or on very thin ($\lesssim 1 \mu$m) layers, for which the scaling is predicted and observed to be $t^{1/2}$ \cite{Troian-1990-MFI, Hamraoui-2004-FPD}.) In our experiments, we measure spreading exponents which approximately agree with the predicted $1/4$ values. However, the Marangoni ridge initially forms behind the leading edge of the surfactant, near the maximum in surfactant concentration, and only later catches up. Thus, the measured exponents are slightly higher for the position of the Marangoni ridge and slightly lower for the position of the leading edge. In the work of Dussaud et al.\ \cite{Dussaud-2005-SCI}, similar behavior was observed, namely a higher spreading exponent for the Marangoni ridge than for the surfactant leading edge. However, some care needs to be taken in connecting these two results, since the ridge position in \cite{Dussaud-2005-SCI} was tracked via the point of minimum surface slope rather than the peak. For a dispersive wave, as is the case here, the peak and the inflection point will become further apart over time. Similarly, the fluorospheres used to visualize the surface disturbance in Bull et al.\ \cite{Bull-1999-SSS} will be moved by the front of the wave associated with the surface compression disturbance, even before the peak of the Marangoni ridge arrives. By directly imaging both the full surface profile and the spatiotemporal evolution of the surfactant concentration, we avoid such difficulties of interpretation.

\section{Experiments} %=========================================================

\begin{figure}
\centerline{\includegraphics[width=5in]{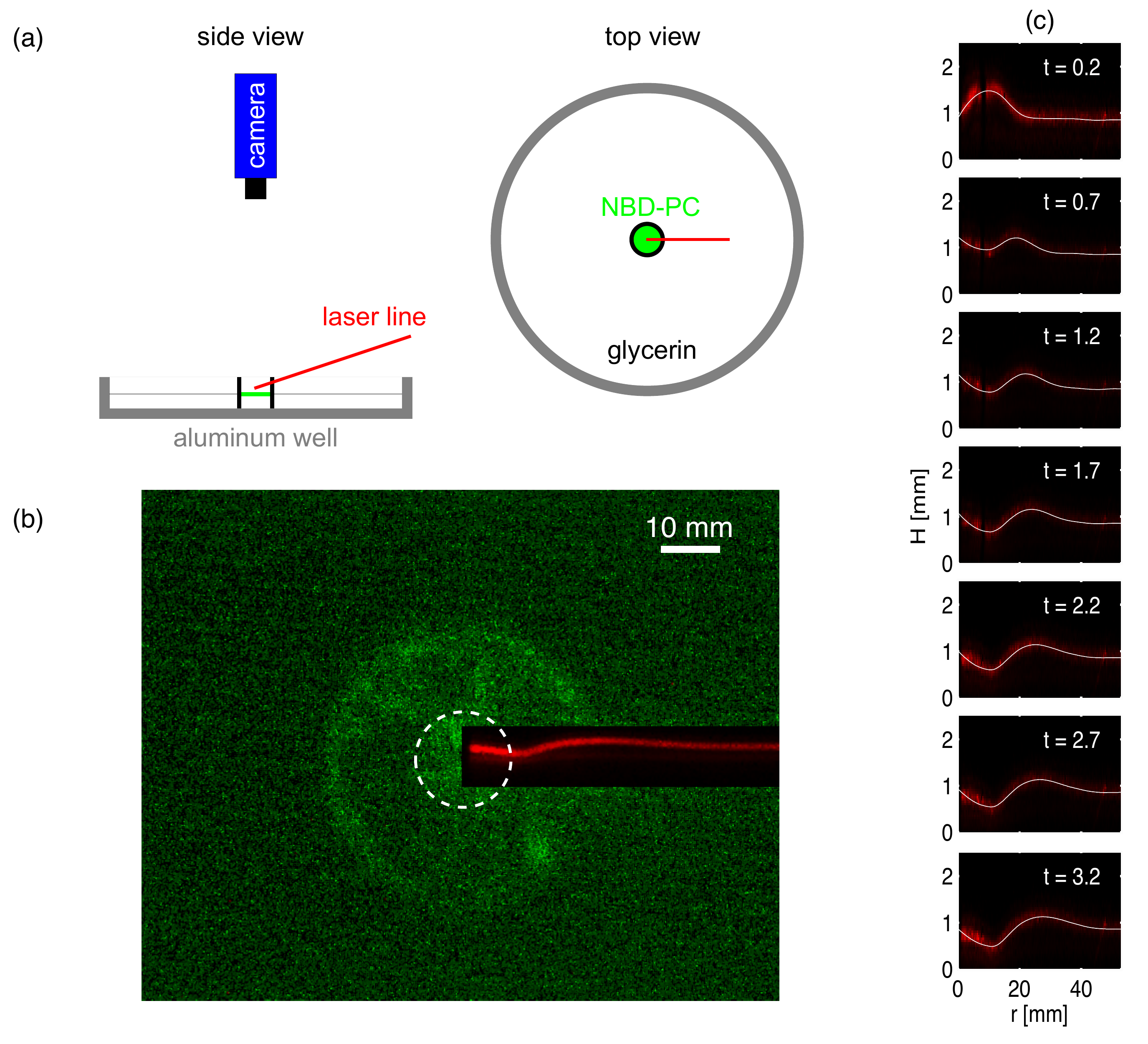}}
\caption{(a) Schematic of initial configuration of experimental apparatus, side and top view. (b) Image of red laser profile of surface height and green fluorescence from surfactant at $t = 2$~sec, with green channel digitally enhanced $\approx40\times$ and white dashed circle marking initial location of confining ring (exposure time 1 second).  The black rectangle denotes the area excluded from fluorescence measurements due to artifacts from the high intensity of the laser signal. See {\tt http://nile.physics.ncsu.edu/pub/movies/thinfilms/} for a movie version of this figure. (c) Time-evolution of laser profile (image) with extracted profile $H(r)$ (white line), for exposure time 0.1 sec.}
\label{setup}
\end{figure}

The experimental cell (see Figure~\ref{setup}a) consists of a machined aluminum well of radius $L = 111$~mm, filled with $38$~mL of $99.5$\% anhydrous glycerin; this results in a fluid layer of depth $d = 0.98 \pm 0.03$~mm. Millimeter-scale fluid layers are chosen to allow for comparison with theories based on the lubrication approximation ($d \ll L$), while avoiding film rupture which would occur for thinner films \cite{Gaver-1992-DST,Dussaud-2005-SCI}. By using viscous (but still Newtonian) glycerin as the working fluid, we are able to achieve sufficiently slow dynamics at low Reynolds number to permit fluorescence measurements of the surfactant concentration.  The glycerin is maintained at a constant temperature of $25 \pm 0.5^\circ$C by circulating water from a temperature-controlled bath through the base of the aluminum well; the viscosity of glycerin at this temperature is $\mu = 0.83 \pm 0.03$~Pa$\cdot$s \cite{Segur-1951}. The surface tension of pure glycerin at this temperature is $\sigma_0 = 63.4$~dyne/cm \cite{Wulf-1999}.  The surface diffusivity of such molecules is small, approximately $10^{-4}$~cm$^2$/s \cite{Sakata-1969-SDM}, and is frequently neglected in modeling the dynamics. Key material parameters and dimensionless ratios are summarized in Table~\ref{paramtable}.

\begin{table}
\begin{small}
\begin{tabular}{rcl} \hline
fluid thickness & $d$ & 0.98 mm \\ \hline
lateral dimension & $L$ & 0.8 cm (ring radius) to 5 cm (final radius) \\ \hline
aspect ratio & $\epsilon \equiv d/L$ &  0.12 (initial) to 0.02 (final) \\ \hline
fluid density & $\rho$ & 1.26 g/cm$^3$ \\ \hline
dynamic viscosity & $\mu$ & 0.83 Pa$\cdot$s \cite{Segur-1951} \\ \hline
bare glycerin surface tension & $\sigma_0$ & 63.4 dyne/cm \cite{Wulf-1999} \\ \hline
surfactant-contaminated surface tension & $\sigma_{m}$ & 35.5 dynes/cm \cite{Bull-1999-SSS}  \\ \hline
spreading parameter & $S \equiv \sigma_0 - \sigma_m$ &  27.9 dynes/cm  \\ \hline
critical monolayer concentration & $\Gamma_{C}$ &  $\approx 0.3$ $\mu$g/cm$^2$ \cite{Bull-1999-SSS}  \\ \hline
surface diffusivity & $D$ & $10^{-4}$~cm$^2$/s \cite{Sakata-1969-SDM} \\ \hline
Bond number (gravity/Marangoni forces)& Bo $ = \rho g h^2/S$ & 0.44 (0.16 for $d = 0.59$ mm)  \\ \hline
capillary/Marangoni forces& $\epsilon^2 \sigma_0/S$ & 0.036 (initial) to 0.00091 (final)  \\ \hline
initial velocity & $v$ & 6 cm/s (maximum measured) \\ \hline
Reynolds number & ${\mathrm Re} = \rho v L / \mu$  & 0.73 (maximum measured) \\ \hline
\end{tabular}
\end{small}
\caption{Summary of key dimensional and dimensionless parameters.  \label{paramtable}}
\end{table}

The spreading surfactant droplet is a fluorescently-tagged lipid from Avanti Polar Lipids, 1-palmitoyl-2-{12-[(7-nitro-2-1,3-benzoxadiazol-4-yl)amino]lauroyl}-sn-glycero-3-phosphocholine (abbreviated NBD-PC). This molecule is a two-tailed lipid with a hydrophilic (also glycerin-philic) head and hydrophobic 16-carbon and 12-carbon tails; the NBD fluorophore is attached to the shorter of the two tails which permits fluorescent visualization. NBD-PC is insoluble in glycerin, lowers the surface tension by as much as 30 dyne/cm \cite{Bull-1999-SSS}, has an absorption peak at 460 nm (excited by a broad-spectrum black light), and has an emission peak at 534 nm. 

The hydrophilic phosphocholine headgroup has a cross-section of approximately $50$~\AA$^2$ \cite{Wilchut-1991} in an
anhydrous medium; from this value, we estimate that $\approx 0.1$~mg of NBD-PC are sufficient to cover the full glycerin surface with a monolayer of molecules. This quantity corresponds to a surfactant monolayer concentration of $\Gamma_{C} \approx 0.3$~$\mu$g/cm$^2$ and represents an upper limit on our monolayer concentration $\Gamma_{C}$ because our liquid layer is not completely anhydrous. For phosphocholines in the presence of water, the cross-section would be no larger than $60$~\AA$^2$ \cite{Wilchut-1991}. This critical concentration $\Gamma_c$ is important because the presence of a larger amount of surfactant will not further reduce the surface tension (i.e., $\Delta\sigma \approx 0$ for $\Gamma > \Gamma_c$). This critical monolayer concentration is in good agreement with measurements by Bull et al.\ \cite{Bull-1999-SSS}, where the equation of state $\sigma(\Gamma)$ becomes flat for $\Gamma \gtrsim 0.3$~$\mu$g/cm$^2$. We conduct our experiments with concentrations which provide partial coverage of the full fluid surface. Most runs use $m = 18~\mu$L of NBD-PC/chloroform solution which contains $18~\mu$g of NBD-PC. This amount corresponds to $1.3\times 10^{16}$ molecules, which would provide a monolayer area with radius $\sim 45$~mm for a headgroup size of $50$~\AA$^2$. The experiments generally spread to approximately this area before we lose resolution in our fluorescence measurements.

To perform surfactant-spreading experiments, we deposit the surfactant within a $16$~mm diameter steel ring ($4$~mm tall) placed in the center of the glycerin layer. Prior to each experimental run the ring and the aluminum well were cleaned with Contrad soap and water then dried using nitrogen gas. The surfactant is initially dissolved in chloroform; we allow at least $6$~minutes of evaporation for each $\mu$L of the solution before lifting the ring to allow the lipid molecules to spread. The surfactant remains confined inside the ring during evaporation, and $18~\mu$g are enough to provide approximately $30$ monolayers within the ring. At the start of each experimental run, we lift the ring using a monofilament line connected by an overhead pulley to a slow motor to provide repeatable dynamics and minimize the formation of bubbles during release.

Our laser profilometer is constructed from a $632.8$ nm HeNe laser fitted with a line generator. We project a straight line of light across the glycerin surface at an angle of $12 \pm 0.5^\circ$ above the horizontal. To create a thin line, the laser light passes through a $0.4$~mm slit; this provides a $2$~mm wide line at the surface of the glycerin. This line is refracted by the glycerin and reflected off the bottom surface of the aluminum well before being imaged by the camera overhead, providing a projection of the height profile of the glycerin surface (see Figure~\ref{setup}b,c). We calibrate the laser deflection as a function of known displacements by noting the change in location of the laser line before and after adding a known thickness of glycerin to the well. Displacement of the laser line is a linear function of the surface displacement.

We make simultaneous measurements of the fluid surface profile and the spatial surfactant distribution via laser profilometry and fluorescence, respectively. Using a color digital camera with a $525$~nm high-pass filter, we can observe both the $632.8$~nm laser profile and the $534$~nm NBD-PC fluorescence while filtering out the UV light used for excitation. The height profile is recorded in the red channel of the camera and the surfactant fluorescence in the green channel, without the need to align two images. We utilize two different frame rates (exposure times) in collecting data in order to optimize either the fluorescence measurements or the fast dynamics. Using a 1 second exposure time, we are able to collect more emitted photons from the fluorescent surfactant and thus obtain improved spatial resolution. Using a 0.1 second exposure time, we are able to obtain greater resolution and earlier times in tracking the position of the Marangoni ridge.

To determine the relationship between fluorescence intensities measured by the camera and surfactant concentration, we take pictures in which known quantities of NBD-PC are placed in 20 mm diameter rings under normal experimental conditions.  The choice of ring size is guided by a desire to minimize errors due to heterogeneity in NBD-PC concentration, yet still provide a large enough area for measurement. The intensities are averaged over a 12 mm diameter region, to avoid shadows at the edges of the ring. Over the range  0.14 to 8.8 $\mu$g/cm$^2$, we find that there is a linear relationship between intensity and NBD-PC concentration. We use this empirical linear relationship to convert fluorescence intensities to surfactant concentrations. Statistical errors in $\Gamma$ are about $\pm 0.05$~$\mu$g/cm$^2$ in the vicinity of the leading edge, and larger near the center of the droplet.

\section{Fluid and Surfactant Dynamics}

In long-exposure images (1 second, as shown in Figure~\ref{setup}b), we can observe both the height profile and the surfactant concentration. Brighter green corresponds to greater surfactant concentration; the red line is a refracted laser line showing the height profile of the glycerin surface. Because the fluorescence signal is weak, we use image-division to remove background heterogeneities and a median filter to eliminate the noisiest pixels. The Marangoni ridge, created by the spreading surfactant, is visible as the peak of the red laser line. The dashed circle indicates the location of the confining ring for the initial surfactant deposit. At early stages, as shown in Figure~\ref{setup}b-f, the spreading surfactant typically has a maximum concentration which trails the leading edge. This maximum is not present in the models, which predict a monotonic decrease \cite{Jensen-1992-ISS, Jensen-1994-SSS, Jensen-1993-SHS, Jensen-1998-SSS1, Jensen-1998-SSS2}. 

Figure~\ref{setup}c shows examples of the raw laser line and the extracted height profile $H(r)$ for short-exposure images (0.1 second). To determine the profile, we find the centroid of pixel intensity for each image column in the red channel of the image, apply a median filter with respect to neighboring columns, and smooth the resulting function using a local linear regression. Within each $H(r)$ we fit a parabola to the Marangoni ridge to determine the location of the peak $R_H$. As the ridge moves outward from the center ($r=0$) over time, the height of the peak decreases due to gravity and capillarity. A trough behind the ridge also develops but does not rupture \cite{Jensen-1992-ISS, Dussaud-2005-SCI} the fluid layer for $d \gtrsim 0.5$~mm. Eventually the peak moves out of the viewing area or becomes indistinguishable from the rest of the glycerin layer. As would be expected, the maximum height is observed to be less than twice the thickness of the fluid layer, the value predicted by models neglecting gravity and capillarity \cite{Jensen-1992-ISS, Jensen-1994-SSS, Jensen-1993-SHS}, and is consistent with models that include gravity \cite{Gaver-1990-DLS}.

\begin{figure}
\centerline{\includegraphics[width=\textwidth]{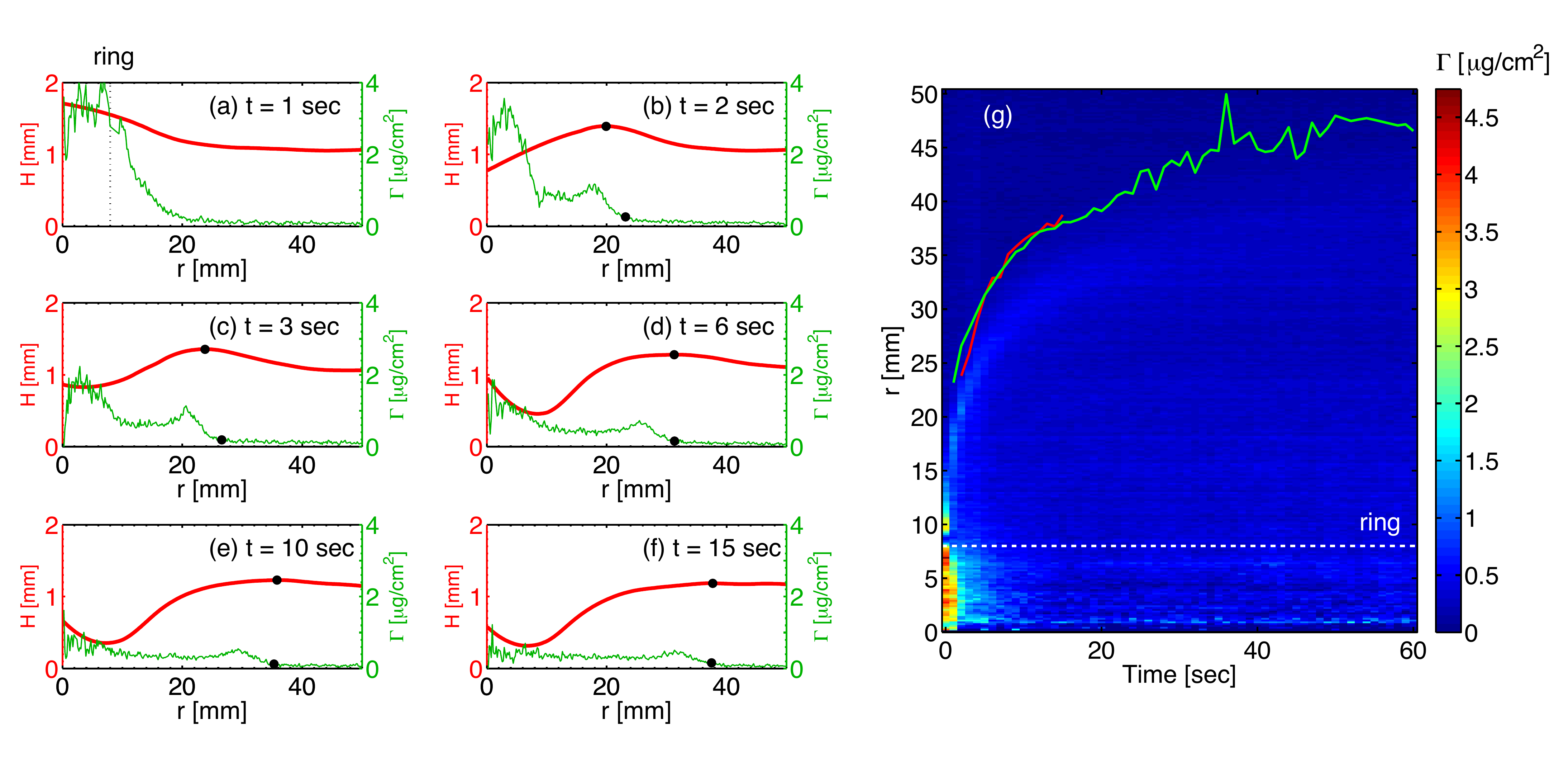}}
\caption{(a-f) Comparison of fluorescence concentration ($\Gamma(r)$, thin green, right axis) and height ($H(r)$, thick red, left axis) profiles. The locations of Marangoni ridge peak $R_H$ and surfactant leading edge $R_\Gamma$ are marked with $\circ$. See {\tt http://nile.physics.ncsu.edu/pub/movies/thinfilms/} for a movie version of this figure. (g) Space-time plot of fluorescence concentration, overlaid with Marangoni ridge peak (solid red line), surfactant leading edge (solid green line), and initial ring radius (horizontal dashed black line) locations.}
\label{spacetime}
\end{figure}

To visualize the surfactant driving this motion, we use the long-exposure images (1 second) and radially average the intensity of the green channel for use in the empirically-measured conversion to concentration. Figure~\ref{spacetime}a-f shows corresponding fluorescence concentration profiles $\Gamma(r)$ and height profiles $H(r)$ at representative times. Figure~\ref{spacetime}a shows the first image after the ring lifts off the surface, with the surfactant still localized to the original ring location (and the shadow of the ring still visible). We set $t=0$ at the middle of the interval captured by the preceding image. As time progresses, a Marangoni ridge develops behind the leading edge of the surfactant and eventually catches up. Note that the fluorescence concentration within the spreading front has a plateau with a $\Gamma \sim \Gamma_c$.  Because the film has not ruptured, the surfactant does not become trapped at the point of deposition as in Dussaud et al.\ \cite{Dussaud-2005-SCI}, and the fluorescence concentration at the center continues to diminish as the surfactant spreads. Importantly, the measured shape of $\Gamma(r)$ differs significantly from what has previously been predicted by various models of similar configurations \cite{Gaver-1990-DLS, Jensen-1992-ISS, Jensen-1993-SHS, Jensen-1994-SSS, Shen-1994-EIC, Jensen-1998-SSS1, Jensen-1998-SSS2, Dussaud-2005-SCI}. In models, $\Gamma(r)$ falls off monotonically without the presence of either the plateau or the leading edge peak observed here.

The distinct peak in surfactant concentration just behind the leading edge of $\Gamma(r)$ is generated during the interval between the first two images (see Fig.~\ref{spacetime}a-b) and might therefore be associated with the detachment of the fluid from the ring. However, in control experiments in which the confining ring is lifted from a uniform layer of surfactant no such peak is formed. Because the surface slope is shallow ($\lesssim 2^\circ$) in the vicinity of the peak, the increased intensity is unlikely to be due to an increased surface area arising from the tilt. The magnitude of the peak is not consistent between runs, and it also appears to be unstable since its magnitude decays in time (see Fig.~\ref{spacetime}b-f). Further work is necessary to understand how it arises and decays, and whether it can be reproduced in models.

An additional unexpected feature is visible in Fig.~\ref{setup}b: spatial heterogeneities in the surfactant concentration which appear as areas of brighter green between the initial ring location and the surfactant leading edge. These heterogeneities form during the initial deposit of the surfactant-chloroform mixture within the ring and do not dissipate due to the low surface diffusivity. The origin of the aggregation likely lies in the electrostatic attractions between the lipid molecules \cite{McLaughlin-1989-EPM}. No characteristic length scale has been observed, but fingering instabilities in spreading surfactant fronts have been observed for flows down inclined planes \cite{Edmonstone-2005-CIP} and for microscopic films \cite{Bardon-1996-SDL, Matar-1997-LSA, Matar-1999-DTF, Cachile-2002-FDS}. Importantly, these heterogeneities likely have only a small effect on the spreading behavior since the spreading front is observed to remain circular at all times. This can be understood as arising from the fact that for $\Gamma > \Gamma_c$, surface tension gradients are small.  Additional work is needed to characterize this  aggregation phenomenon, for which the ability to directly observe the surfactant location is a key technique.

To visualize the spatial relationship between the Marangoni ridge and the surfactant concentration, Figure~\ref{spacetime}g shows the positions of the peak of $H(r)$ and the leading edge of $\Gamma(r)$ with respect to the underlying fluorescence concentration field $\Gamma(r,t)$. At early times ($t \lesssim 6$~seconds), the Marangoni ridge peak lags behind the surfactant leading edge. After about 6~seconds, the peak of the Marangoni ridge catches up to the leading edge of the surfactant, which corresponds to the behavior observed in \cite{Jensen-1992-ISS}. The peak and leading edge move together for the remaining time we are able to track the position of the Marangoni ridge peak, which becomes too flat to detect after about 20 seconds. Therefore, no long-time comparison of the two positions is possible.

\begin{figure}
\centerline{\includegraphics[width=6in]{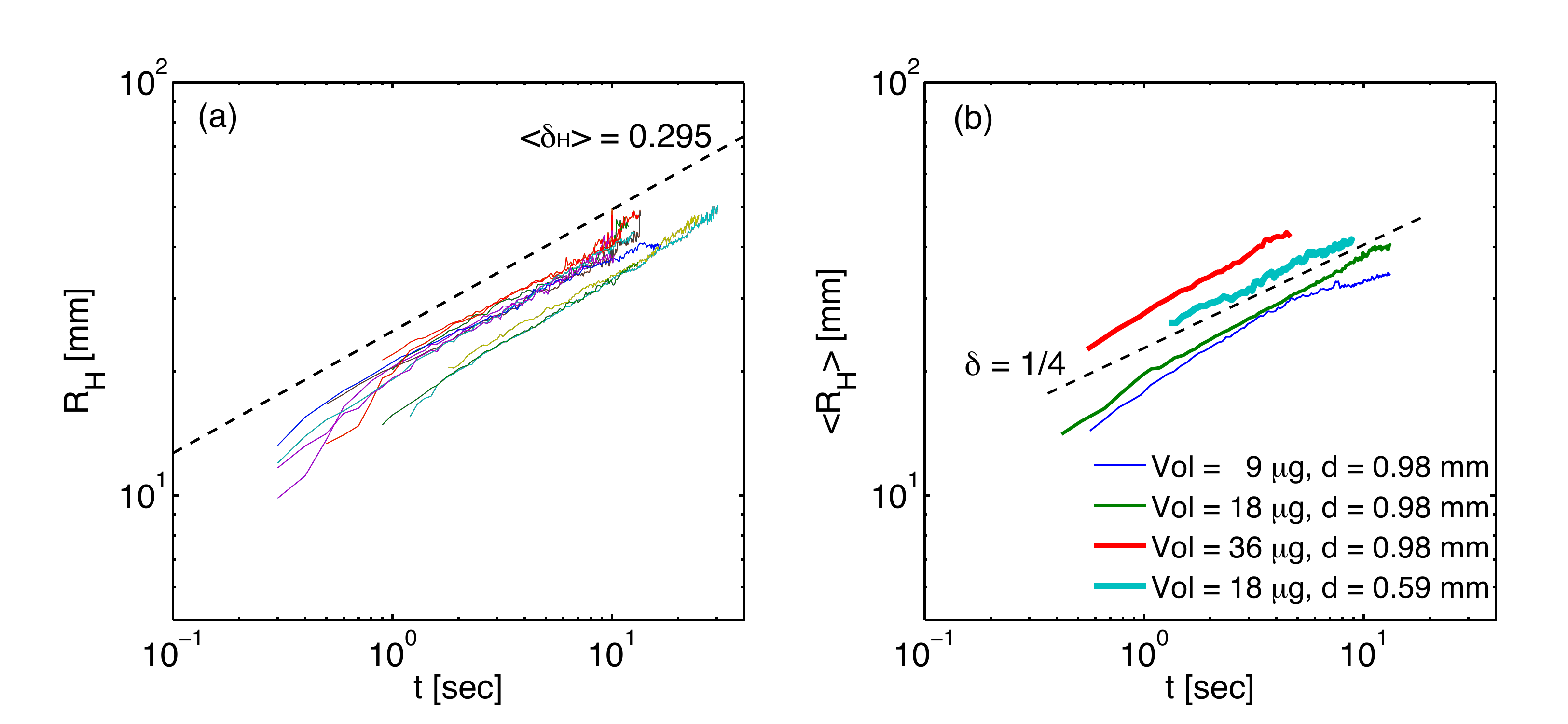}}
\caption{(a) Marangoni ridge location for eleven runs with $m=18 \, \mu$g (thin solid lines) and $R_H(t) \propto t^{\langle\delta_H\rangle}$ for the mean spreading exponent $\delta_H$ measured for each run (dashed line). (b) Comparison of $\langle R_H(t)\rangle$ for several combinations of $m$ and $d$:  $m= 9 \, \mu$g and $d = 0.98$~mm (3 runs), $18 \, \mu$g  and $d = 0.98$~mm (11 runs shown in (a)),  $36 \, \mu$  and $d = 0.98$~mm (2 runs), and $m = 18 \, \mu$g and $d = 0.59$~mm (2 runs). Dashed black line shows comparison to spreading exponent $\delta=1/4$.}
\label{peaktraj}
\end{figure}

To quantify the spreading of both the Marangoni ridge and the surfactant, we examine the scaling of the Marangoni ridge position $R_H(t)$ and the leading edge position $R_\Gamma(t)$. Figure~\ref{peaktraj}a shows $R_H(t)$ for 11 experimental runs with $m = 18 \, \mu$g of NBD-PC, with each line representing a different experimental run. There is considerable scatter in the initial $R_H(0)$, but the exponent is consistent between runs. We fit each curve to the form $R_H \propto t^{\delta_H}$ and find a mean spreading exponent of $\langle \delta_H \rangle = 0.295 \pm 0.03$, represented by the slope of the dashed line. 

We examine the role of surfactant concentration by comparing runs with initial mass $m$ of 9, 18 and 36~$\mu$g. Figure~\ref{peaktraj}b shows $\langle R_H(t) \rangle$ for each case, where the average is taken over the ensemble of runs. Interestingly, the spreading behaviors for $m=9$ and 18~$\mu$g are very similar, while the 36~$\mu$g has a larger prefactor, but still the same exponent $\delta_H$. It is possible that the faster rate is related to reservoir effects stemming from the additional supply of molecules, but in all cases the initial supply concentration exceeds one monolayer. For a thinner layer ($d = 0.59$~mm) which better-satisfies the conditions of the lubrication approximation without being so thin as to cause surface rupture, we observe a spreading exponent closer to $1/4$.

\begin{figure}
\centerline{\includegraphics[width=3in]{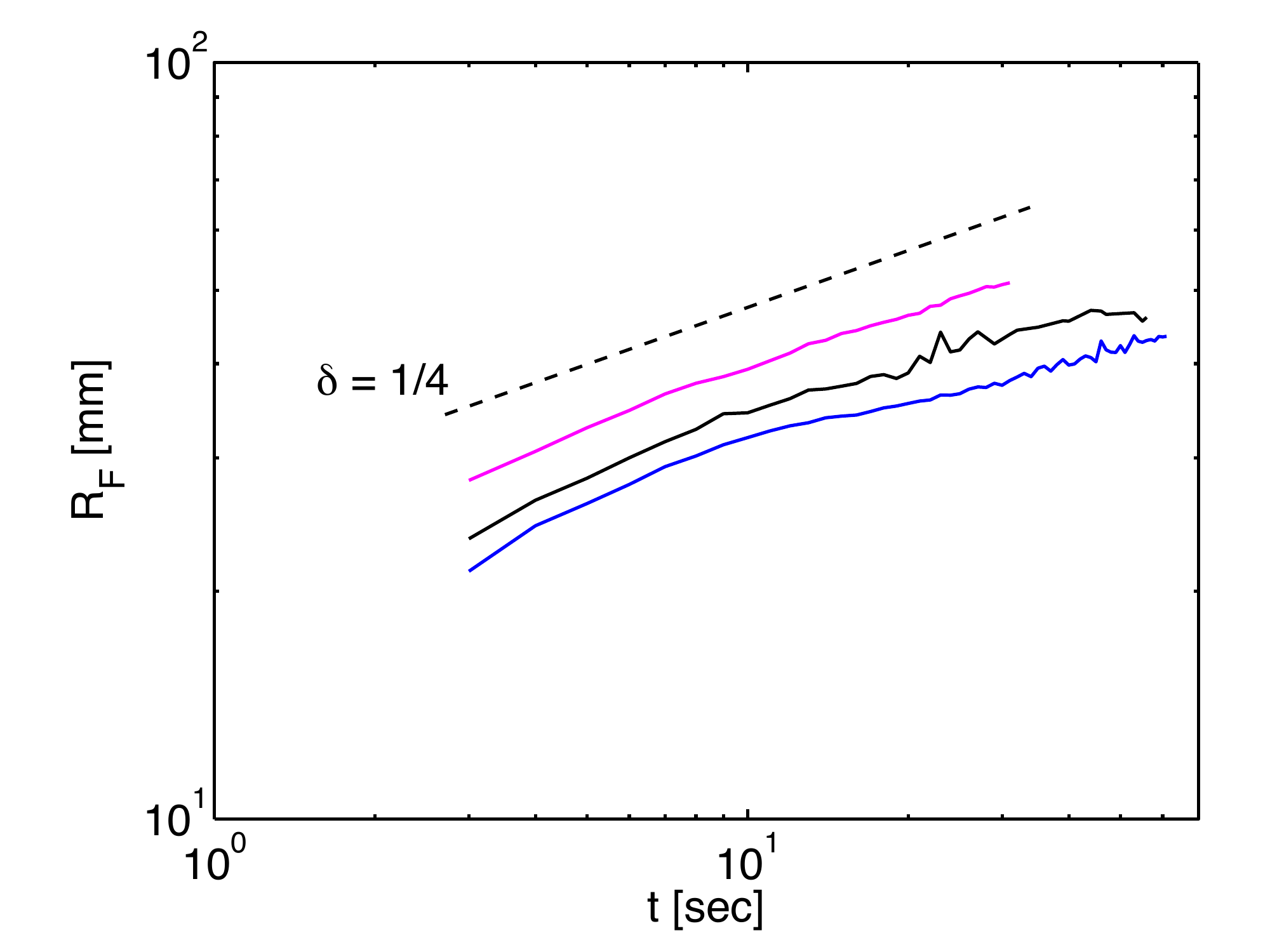}}
\caption{(a) Position of the surfactant leading edge for three runs with $m=18 \, \mu$g. The dashed black line shows a comparison to spreading exponent $\delta_\Gamma=1/4$. }
\label{leadsurf}
\end{figure}

Figure~\ref{leadsurf} shows a similar plot for $R_\Gamma(t) \propto t^{\delta_\Gamma}$, using images with 1 second exposure times. We automatically located the leading edge by finding this peak in the intensity gradient of a smoothed profile, fitting a parabola to the peak, and taking the edge to be the position at which the parabola crosses zero. Our ability to detect the edge is limited at long times by the decreasing intensity of the signal. However, we do not see a change in behavior as the system approaches the estimated $45$~mm radius at which the NBD-PC molecules become a monolayer. Figure~\ref{spacetime}f supports the identification of this length scale, as the surfactant concentration profile $\Gamma(r)$ has become nearly flat behind the leading edge at $R_\Gamma = 40$~mm. Three separate runs with $m=18 \, \mu$g  yield $\langle \delta_\Gamma \rangle \approx 0.22 \pm 0.03$, which is slower than observed for the Marangoni ridge position, and in approximate agreement with the $1/4$ scaling predicted by Jensen and Grotberg \cite{Jensen-1992-ISS, Jensen-1994-SSS}.

\section{Discussion and Conclusions}

The combined techniques of laser profilometry and fluorescent surfactants have allowed us to simultaneously monitor both the surface deformation of a glycerin layer and the spreading of surfactant molecules across the surface. Releasing a mass of surfactant from within a confining ring allows us to compare our experimental observations with models and simulations for the spreading of an axisymmetric drop as in \cite{Jensen-1992-ISS, Jensen-1994-SSS} which predict $R \propto t^{1/4}$. We find that the measured exponents $\langle \delta_\Gamma \rangle = 0.22 \pm 0.03$ and $\langle \delta_H \rangle  = 0.295 \pm 0.03$ are both in reasonable agreement with the predicted exponent. It is important to note that the exponent $1/4$ is in fact only expected under a set of highly restrictive conditions which neglect gravity, neglect capillarity, and utilize the lubrication approximation ($\epsilon \ll 1$). Furthermore, the lifting of the ring creates a disturbance with its own length scale: it is therefore likely that no single exponent fully characterizes the spreading. As observed in the models, the peak of the Marangoni ridge and the location of the surfactant leading edge coincide (although only after an initial transient during which the ridge lags). In addition, the fluid profile is observed, as in \cite{Gaver-1990-DLS}, to have a capillary trough which trails the Marangoni ridge. The fluorescence imaging indicates that this trough is associated with a reservoir for the spreading surfactant. 

Interestingly, the spreading exponent $\delta_H$ (measured for the peak of the Marangoni ridge) is higher than $\delta = 1/4$ but $\delta_\Gamma$ (obtained from fluorescence measurements of the position of the leading edge of the surfactant) is lower. While the range of times over which we can take measurements of $R_{H}$ and $R_{F}$ are overlapping, early-time data is only available for $R_{H}$ and late-time data is only available for $R_{F}$. Since $R_\Gamma$ is measured over both longer and later times than $R_H$, $\delta_\Gamma$ would be expected to be closer to the asymptotic result. The higher-than-expected value of $\delta_H$ may simply reflect the lingering effects of the initial disturbance provided by the ring lifting and then detaching from the fluid surface. Thus, its possible that an (unobserved) crossover to $\delta_\Gamma = \delta_H$ occurs for later times than were measured.

In addition, several other discrepancies from the idealized model \cite{Jensen-1992-ISS, Jensen-1994-SSS} might act separately on the two positions $R_H$ and $R_\Gamma$. Capillarity and gravity do not act on the fluid and surfactant in the same way. Second, while the ratio of film thickness to lateral extent is small, three-dimensional effects (such as circulation under the free surface) are likely still present and may affect the spreading rate of the fluid (and to a lesser extent, the surfactant). Indeed, Gaver and Grotberg \cite{Gaver-1992-DST} calculated that there is a re-circulation of particle-trajectories within the bulk of a fluid, but it remains unclear what consequence this has on the Marangoni ridge motion. Further efforts to probe the validity of the lubrication approximation in this regime are necessary.

To understand the low value of $\delta_\Gamma$, we should consider the surfactant itself. 
In our initial configuration, the confining ring contains multiple layers of surfactant. These multiple layers may act as a reservoir of surfactant and thus might provide an effective flux of surfactant. However, the predicted effect \cite{Jensen-1992-ISS} is $\delta = (1 + \alpha)/4$ for a constant flux of surfactant supplied at a rate of $t^\alpha$, which would necessarily lead to a larger value for $\delta_\Gamma$, rather than the smaller value observed. In addition, we do not observe a change in $\delta_H \approx 0.3$ even when we change the initial number of monolayers via initial surfactant mass $m$. A second effect is the presence of a non-zero surface diffusivity \cite{Sakata-1969-SDM}, which is predicted to increase $\delta$ by amounts similar to the discrepancies observed here \cite{Gaver-1990-DLS}. However, this would shift $\delta_{\Gamma}$ in the opposite direction from what is observed.

Significantly, two key discrepancies with existing models \cite{Gaver-1990-DLS, Jensen-1992-ISS, Jensen-1993-SHS, Jensen-1994-SSS, Shen-1994-EIC, Bull-1999-SSS, Dussaud-2005-SCI} arise in examining the shape of the surfactant concentration $\Gamma(r)$: neither the presence of a plateau behind the leading edge nor the peak immediately behind the leading edge are predicted. Instead, the model $\Gamma(r)$ falls off monotonically. Interestingly, the plateau has a near-monolayer concentration ($\Gamma \sim \Gamma_c$). At early times, the region behind the leading edge has $\Gamma > \Gamma_c$, and therefore only small surface-tension gradients would act to smooth the peak in $\Gamma$. Thus, there is no observable disturbance to $H(r)$ due to peak in $\Gamma(r)$. Further modeling work will be required in order to understand how both the extended plateau and the peak arise and ultimately decay.

\bigskip 

The authors wish to thank Michael Shearer, Rachel Levy, Ellen Peterson, and Omar Matar for insightful discussions during the development and analysis of these experiments, as well as three anonymous reviewers. Keith Weninger provided a spectrophotometer for use in measuring the emission and absorption peaks of the NBD-PC. Our research was supported by the NSF under grants DMS-0604047 (DWF, AML, and KED) and DMR-0353719 (CJF).

\section*{References}

%\bibliographystyle{unsrt} 
%\bibliography{ked,dave}

\end{document}